\documentclass[prd,aps,superscriptaddress,showpacs,preprintnumbers,amsmath,amssymb,floatfix,12pt]{revtex4}

\usepackage{graphicx}
\usepackage{feynmp}
\preprint{CU-TP-1131, RBRC-530, SHEP-0520}
\begin{document}
\bibliographystyle{apsrev}

\title{Finite Volume Corrections to the Two-Particle Decay of States 
with Non-Zero Momentum}

\author{Norman H.~Christ}
\affiliation{Physics Department,Columbia University,New York, NY 10027}

\author{Changhoan Kim}
\affiliation{Physics Department,Columbia University,New York, NY 10027}
\affiliation{School of Physics and Astronomy, University of Southampton,
Southampton, SO17 1BJ, UK.}

\author{Takeshi Yamazaki}
\affiliation{RIKEN-BNL Research Center, Brookhaven National Laboratory,Upton, NY 11973}

\bibliographystyle{apsrev}

\begin{abstract}
We study the effects of finite volume on the two-particle decay rate
of an unstable state with non-zero momentum.  First L\"uscher's 
field-theoretic relation between the infinite volume scattering  phase 
shifts and the quantized energy levels of a finite volume, two-particle
system is generalized to the case of non-zero total momentum and compared 
with the earlier results of Rummukainen and Gottlieb.  We then use 
this result and the method of Lellouch and L\"uscher to determine the
corrections needed for a finite-volume calculation of a two-particle 
decay amplitude when the decaying particle has non-vanishing 
center-of-mass momentum.
\end{abstract}

\maketitle

\newpage

\section{Introduction}
\label{sec:introduction}

A central problem in making Standard Model predictions for $K$ meson
decays, including the important CP violating amplitudes, is  
calculation of the decay $K \rightarrow \pi\pi$ into the $I=0$ final state 
of two pions.  In a standard calculation in which the $K$ meson is at
rest, the contribution of the two pion final state is expected to be 
very difficult to extract from a lattice QCD calculation.  The physical 
state has the same quantum numbers as the state with the two pions at rest
(the difficulty emphasized by Maiani and Testa \cite{Maiani:1990ca}).
Of even greater concern is the fact that the QCD vacuum also has these
same quantum numbers.  A very appealing approach to deal with these
problems is to study the decay of a $K$ meson with non-zero momentum.
For a calculation in finite volume, it is possible to adjust the
momentum of the $K$ meson and the box size so that both the transition 
amplitude to the vacuum must vanish (since the vacuum has zero momentum) 
and the final two-pion state must have physical relative momentum
and total energy equal to that of the decaying K meson.

For such a calculation to be useful we must be able to relate 
the decay amplitudes computed in finite volume using this technique to
the infinite volume matrix elements determined by experiment.  To a
large extent the effects of finite volume on the particles involved
are relatively mild.  Typically the volumes used in such a 
calculation can be chosen sufficiently large that they do not 
significantly distort the $K$ and $\pi$ particles whose physical size
will be much smaller than that of the box employed.  The most obvious
finite-volume effect will be the quantization of the energy levels of 
the two-pion final state.  This is actually an advantage, permitting
the box size to be adjusted to make one of the discrete, two-pion
energies match precisely that of the $K$ meson.  For the case at
hand, this can even be the lowest energy $\pi-\pi$ state.  

However, the violation of rotational symmetry by the finite-volume 
boundary conditions does induce an important distortion in the 
computed decay rate.  Because of the resulting non-conservation
of angular momentum, the two-particle state into which the decay
occurs is actually a mixture of states with many angular momenta.
For typical lattice volumes the actual decay of the $K$ meson
into these higher angular momentum states will be very small---angular
momentum is effectively conserved at the short distances over
which the decay occurs.  However, the presence of these extra
angular momentum states effects the normalization of the physical 
$J=0$ amplitude which appears in the matrix element.

This finite volume normalization problem has been solved by Lellouch 
and L\"uscher \cite{Lellouch:2000pv} for the case of the decay of a
$K$ meson at rest.  This problem has also been solved by a different
approach in Ref.~\cite{Lin:2001ek}.  In this paper we will generalize 
the approach of Lellouch and L\"uscher to obtain a result for states with 
non-zero total momentum.  Central to their argument is an earlier treatment 
of L\"uscher \cite{Luscher:1986pf,Luscher:1990ux} which determines the 
allowed, finite-volume, two-particle energy eigenvalues in terms of 
the infinite volume, two-particle scattering phase shifts for 
energies below all inelastic thresholds.  This discussion must also 
be generalized to the case of non-zero center-of-mass momentum.

This topic has been studied earlier by Rummukainen and Gottlieb 
\cite{Rummukainen:1995vs}.  Their treatment involves an application
of relativistic two-particle quantum mechanics.  We believe that
it is also important to study this problem starting from the
equations of quantum field theory.

Our strategy follows closely that of L\"uscher \cite{Luscher:1986pf,
Luscher:1990ux} and Lellouch and L\"uscher \cite{Lellouch:2000pv}.   
We first discuss the energy quantization of finite-volume, interacting, 
two-particle states with non-zero center-of-mass momentum.  Following 
L\"uscher, this is first done in standard, two-particle, non-relativistic 
quantum mechanics in Sec.~\ref{sec:2-part-non-rel}.  In 
Sec.~\ref{sec:2-part-rel} we begin with the Bethe-Salpeter equation of 
relativistic field theory and, again following L\"uscher, show how this 
equation when restricted to a particular 7-dimensional subspace of the 
8-dimensional, 2-particle momentum space reduces to the standard 
Lippmann-Schwinger equation describing the earlier non-relativistic 
system.  Finally, in Sec~\ref{sec:decay} we use this result to generalize 
the argument of Lellouch and L\"uscher to the determine the finite volume 
corrections to the decay amplitude computed for states with non-zero 
total momentum.

The issues addressed in this paper have also been considered by Kim, 
Sachrajda and Sharpe. Using a related but different approach, they 
have also confirmed the validity of the results of Ref.~\cite{Rummukainen:1995vs} 
and derived the generalization of the result of Lellouch and L\"uscher 
for the case of non-zero total momentum.  Their paper~\cite{kss} is 
being released simultaneously with the present article.

\section{Finite volume, non-relativistic, two-particle states}
\label{sec:2-part-non-rel}

We begin by considering a simple, non-relativistic system of two
distinguishable particles confined in a cubic box of side $L$ 
and obeying periodic boundary conditions.  The system is described
by a wave function $\psi(\vec r_1, \vec r_2)$ which is periodic
in $\vec r_1$ and $\vec r_2$ separately.  An eigenstate of energy
$\psi_E$ obeys the Schr\"odinger equation:
\begin{equation}
\Bigl\{-\frac{\nabla_1^2}{2m} - \frac{\nabla_2^2}{2m} 
         + V(|\vec r_1 - \vec r_2|)\Bigr\}\psi_E(\vec r_1, \vec r_2) 
       = E \psi_E(\vec r_1, \vec r_2),
\end{equation}
where $m$ is the identical mass of the two particles and 
$V(|\vec r_1 - \vec r_2|)$ their rotationally invariant interaction 
potential.

Such an equation is conventionally simplified by changing to center-of-mass
and relative coordinates: 
\begin{eqnarray}
\vec R &=& \frac{\vec r_1 + \vec r_2}{2} \label{eq:rel_coord1} \\
\vec r &=& \vec r_1 - \vec r_2           \label{eq:rel_coord2}
\end{eqnarray}
with conjugate momenta
\begin{eqnarray}
\vec P &=& \vec p_1 + \vec p_2 \\
\vec p &=& \frac{\vec p_1 - \vec p_2}{2},
\end{eqnarray}
where $\vec p_i$ is the momentum conjugate to the coordinate $\vec r_i$.

Using $\psi_E^{\rm (rel)}(\vec r, \vec R)$ to represent the original wave function
expressed in terms of $\vec r$ and $\vec R$, we can write down the 
standard equation which it obeys:
\begin{equation}
\Bigl\{-\frac{\nabla_R^2}{4m} - \frac{\nabla_r^2}{m} 
         + V(|\vec r|)\Bigr\}\psi_E^{\rm (rel)}(\vec r, \vec R) 
       = E \psi_E^{\rm (rel)}(\vec r, \vec R),
\label{eq:non-rel-1p}
\end{equation}
The periodicity of $\psi(\vec r_1, \vec r_2)_E^{\rm (rel)}$ under the 
simultaneous translation $\vec r_i \rightarrow \vec r_i + \hat e_k L$ 
(where $\hat e_k$ is a unit vector parallel to one of the edges of 
the box) implies the periodicity of the wave function $\psi_r$ under a 
translation of $R$ by $L$.  Thus, the conserved total 
momentum $\vec P$ must obey the quantization condition: 
$\vec P = \sum_{k=1}^3 2\pi n_k \hat e_k /L$ for integer $n_k$.  
It is easy to see that in a direction $k$ for which the integer 
$n_k$ is even, the corresponding component of the relative momentum 
$p_k = 2\pi n^\prime_k /L$  while if $n_k$ is odd then 
$p_k = 2\pi(n^\prime_k+\frac{1}{2})/L$, where $n^\prime_k$ is an 
integer.  With this change of coordinates and a specific choice of 
$\vec P$, our two-particle problem reduces to the quantum mechanics of a
single particle in an $L^3$ box obeying either periodic or antiperiodic
boundary conditions on each of its three opposing faces.  

In infinite volume, this can be viewed as a scattering problem often 
phrased as a Lippmann-Schwinger integral equation.  One defines an
energy eigenstate $\psi^{\rm in}_{\vec p}$ whose incoming part 
(that term with radial dependence $e^{-ipr}$) is that of a plane
wave with momentum $\vec p$.  Necessarily, the outgoing part of 
$\psi^{\rm in}_{\vec p}$ is more general, being created by scattering 
from the potential $V$.  Such a state must have energy $E\,^{\rm cm}=p^2/m$.  
Manipulation of Eq.~\ref{eq:non-rel-1p} easily produces the desired 
integral equation:
\begin{equation}
\psi^{\rm in}_{\vec p} = \phi_{\vec p} + 
                 \frac{1}{E\,^{\rm cm} - H_0 +i\epsilon}V\psi^{\rm in}_{\vec p},
\label{eq:lippmann-schwinger}
\end{equation}
where $\phi_{\vec p}$ is the plane wave solution 
$\phi_{\vec p}(\vec r)=e^{i\vec p\cdot\vec r}$ of the free 
Schr\"odinger equation and $H_0 = -\nabla^2/m$, the free Hamiltonian of 
a particle with the ``reduced mass'' $m/2.$  Of course the full solution
to Eq.~\ref{eq:non-rel-1p}, $\psi_E^{\rm (rel)}(\vec r, \vec R)$, is a 
product of a plane-wave depending on the center-of-mass coordinate 
$\vec R$ and the wave function above:
\begin{equation}
\psi_E^{\rm (rel)}(\vec r, \vec R) = e^{i \vec P \cdot \vec R} 
                                      \psi^{\rm in}_{\vec p}(\vec r)
\end{equation}
and the total energy $E$ is related to the energy in the center-of-mass
system by $E = E\,^{\rm cm} + \vec P^2/4m$.

Examining the asymptotic behavior of Eq.~\ref{eq:lippmann-schwinger}, 
one derives the standard relation between the conventional scattering 
amplitude $f(\theta)$ and the matrix element of $V$ between the plane 
wave state $\phi_{\vec p\,^\prime}$ and $\psi^{\rm in}_{\vec p}$:
\begin{equation}
f(\theta) = -2\pi^2 m\langle \phi_{\vec p\,^\prime} |V| \psi^{\rm in}_{\vec p} \rangle,
\end{equation}
where $\vec p\,^\prime \cdot \vec p = p^2 \cos(\theta)$.  We obtain an
equation closer to the relativistic Bethe-Salpeter equation by defining
the $T$ matrix as:
\begin{equation}
\langle \vec p\,^\prime | T | \vec p \rangle 
= \langle \phi_{\vec p\,^\prime} |V|\psi^{\rm in}_{\vec p} \rangle.
\label{eq:T-definition}
\end{equation}
Here we are using the conventional Dirac bra-ket notation to represent
momentum eigenstates: $\phi_{\vec p} \equiv |\vec p \rangle$.  Note, 
Eq.~\ref{eq:T-definition} defines matrix elements of $T$ even when 
$|\vec p\,^\prime| \ne |\vec p|$ and energy is not conserved.  

Using the matrix $T$ we can rewrite Eq.~\ref{eq:lippmann-schwinger} 
in a form that will be useful later, if we multiply by $V$ and transform
to momentum space:
\begin{equation}
\langle \vec p\,^\prime|T|\vec p\rangle 
                 = \langle \vec p\,^\prime|V|\vec p\rangle
                  + \int \frac{d^3 k}{(2\pi)^3} 
                    \langle \vec p\,^\prime|V|\vec k\rangle
                    \frac{1}{p^2/m - k^2/m +i\epsilon}
                    \langle \vec k|T|\vec p \rangle.
\label{eq:b-s-non-rel}
\end{equation}
This equation explicitly involves the matrix elements of $T$ between states 
with different energies.  Equation~\ref{eq:b-s-non-rel} is a component of standard
scattering theory and applies only to the case of infinite volume.  Since we
are interested also in the eigenfunctions and energies for the finite volume
problem it will be helpful to cast the standard Schr\"odinger equation, obeyed
by an eigenstate, $\psi_n(\vec r)$ with the discrete energy $E^{\rm cm}_n$
\begin{equation}
         (H_0+V)\psi_n = E^{\rm cm}_n\psi_n,
\label{eq:se_diff}
\end{equation}
into a similar form:
\begin{eqnarray}
         V\psi_n &=& V\frac{1}{E^{\rm cm}_n-H}V\psi_n \nonumber \\
                     &=& \int \frac{d^3 k}{(2\pi)^3} 
                    \langle \vec p\,^\prime|V|\vec k\rangle
                    \frac{1}{E^{\rm cm} - k^2/m}
                    \langle \vec k|T|\vec p \rangle.
\label{eq:se_intgr}
\end{eqnarray}
This equation demonstrates that the state $V\psi_{E_n}$ solves
the homogenous Lippmann-Schwinger equation if the energy argument in the
denominator, $E^{\rm cm} = \vec p^2/m$ is continued to that state's 
actual energy $E^{\rm cm}=E_n^{\rm cm}$.  Equation~\ref{eq:se_intgr} can be 
used in infinite volume to determine the energy of a possible bound state.
It also can be applied to the case of finite volume if one makes a simple 
replacement of the integral over the relative momentum $\vec k$ by an 
appropriate discrete sum.

In Eq.~\ref{eq:b-s-non-rel} we have followed the standard procedure, 
exploiting the separation of relative and center-of-mass variables 
permitted by Eq.~\ref{eq:non-rel-1p}, and written that integral equation 
as an equation obeyed by functions of a single,
three-dimensional, relative momentum $\vec k$ or $\vec p$.  The three-momentum 
of the center-of-mass, $\vec P$, disappears from the problem once we use the
energy in center-of-mass system, $E\,^{\rm cm}$.  However, we could create a
more explicit analogy with the relativistic discussion to follow if we 
viewed the states $|\vec p \rangle$ and $|\vec k \rangle$ as functions of
the four-momentum of the center-of-mass as well: 
$P(\vec p)= (\frac{\vec P^2}{4m} + \vec p^2/m,\vec P)$ and 
$P(\vec k)= (\frac{\vec P^2}{4m} + \vec k^2/m,\vec P)$ respectively.  Had we
done this, Eq.~\ref{eq:b-s-non-rel} would be an equation obeyed by functions
of seven variables.  Each factor in this equation would be explicitly diagonal 
in $\vec P$ and the difference of these total energy variables $P_0(\vec p)$ 
and $P_0(\vec k)$ would replace the present denominator in that equation.
(A similar remark applies to Eq.~\ref{eq:se_intgr} as well.)

The final step to be reviewed in this section is the connection between the 
infinite volume scattering problem, defined by Eq.~\ref{eq:b-s-non-rel}, and
the finite volume energy eigenvalues of the original Schr\"odinger equation,
Eq.~\ref{eq:non-rel-1p} or Eq.~\ref{eq:se_intgr}.  This is the problem 
solved by L\"uscher in Refs.~\cite{Luscher:1986pf,Luscher:1990ux}.  Recall 
that the scattering amplitude $f(\theta)$ can be written as a sum over partial 
waves as:
\begin{equation}
f(\theta) = \sum_{l=0}^\infty (2l+1)\frac{e^{i 2\delta_l}-1}{2 i p}P_l(cos(\theta)),
\end{equation}
where the $\delta_l$ are the standard scattering phase shifts and the
$P_l(cos(\theta))$ the usual Legendre polynomials.  In 
Refs.~\cite{Luscher:1986pf,Luscher:1990ux}, L\"uscher examines the case of a 
potential of finite range, $V(r) = 0$ for $r > R_{\rm max}$, and for the case 
$L > 2R_{\rm max}$ derives a relation between the allowed energies $E\,^{\rm cm}=p^2/m$ 
in the finite box and the phase shifts $\delta_l$.  For the simplest case where
all $\delta_l \approx 0$ for $l > 0$, this finite-volume quantization condition
becomes:
\begin{equation}
n\pi-\delta_0(p) = \phi(q)
\label{eq:quantize}
\end{equation}
where $n$ is an integer, $p= \sqrt{m E}$, $q = p L/2\pi$ and the function
$\phi(q)$ is a known kinematic function given by:
\begin{equation}
\tan \phi(q) = -\frac{\pi^{3/2}q}{{\cal Z}_{00}(1;q^2)}, \quad \phi(0)=0,
\label{eq:phi-def}
\end{equation}
with the zeta function ${\cal Z}_{00}(s;q^2)$ defined by 
\begin{equation}
{\cal Z}_{00}(s;q^2) = \frac{1}{\sqrt{4\pi}} \sum_{n \in {\mathbb Z}^3}(n^2-q^2)^{-s}.
\label{eq:zeta}
\end{equation}
The zeta function defined above applies to the case that the integers
appearing in the center of mass momentum $\vec P$ are even.  For the case that
one or more is odd, the finite volume problem will obey anti-periodic boundary 
conditions in those directions and an appropriate offset of $1/2$ must be added 
in the summation in Eq.~\ref{eq:zeta}, as discussed in 
Ref.~\cite{Rummukainen:1995vs}.  We conclude that the total energy 
of a finite volume system with total momenta $\vec P$, is given by 
$E = \vec P^2/4m + E^{\rm cm}$ where $p=\sqrt{mE^{\rm cm}}$ obeys 
Eq.~\ref{eq:quantize}.

Thus, L\"uscher's relation for non-relativistic quantum mechanics between the 
scattering phase shifts and the energy eigenvalues for that same system in a 
finite volume is straight forward to generalize to the case that the 
two-particle system carries non-zero total momentum.  We now consider the 
generalization of the next step to non-zero total momentum: the connection 
between the Bethe-Salpeter equation of relativistic quantum field theory and 
Eq.~\ref{eq:b-s-non-rel} above, obeyed by the non-relativistic $T$ matrix.  
This represents the new result of this paper.

\section{Finite volume relativistic two-particle states}
\label{sec:2-part-rel}

In this section we will generalize to the case of non-zero total momentum the 
procedure introduced by L\"uscher to reduce the Bethe-Salpeter equation of 
relativistic field theory to an equation whose form is identical to the 
non-relativistic Eqs.~\ref{eq:b-s-non-rel} and \ref{eq:se_intgr}.  This will 
permit the quantization condition described in Eqs.~\ref{eq:quantize}-\ref{eq:zeta} 
to be applied to quantum field theory, and QCD in particular.  As discussed in 
Refs.~\cite{Luscher:1986pf,Luscher:1990ux}, we must do this by carefully
distinguishing effects of finite volume which fall as powers or exponentially
in the system size.  Note: our objective is to derive an equation which is both 
similar in form to Eqs.~\ref{eq:b-s-non-rel} and \ref{eq:se_intgr} and also 
accurate for both finite and infinite volume so that it can be used 
to relate the finite volume spectrum and the infinite volume scattering 
amplitude.

We begin with the standard Bethe-Salpeter equation, which connects the 
amputated four-point function, $T(p_1^\prime, p_2^\prime;p_1, p_2)$,
the two-particle irreducible kernel $K(p_1^\prime, p_2^\prime;p_1, p_2)$ 
and the single particle propagator, $\Delta(k^2)$:
\begin{eqnarray}
T(p_1^\prime, p_2^\prime;p_1, p_2)
              &=& K(p_1^\prime, p_2^\prime;p_1, p_2) \nonumber \\
              &&+\int \frac{d^4 \bar k}{(2\pi)^4} 
                     K(p_1^\prime, p_2^\prime;P/2+\bar k,P/2-\bar k) \nonumber \\
              &&\cdot\Delta((P/2+\bar k)^2)\Delta((P/2-\bar k)^2) 
                 T(P/2+\bar k,P/2-\bar k,p_1, p_2),
\label{eq:b-s-rel}
\end{eqnarray} 
where $P=p_1+p_2=p_1^\prime+p_2^\prime$ is the total four-momentum.  For simplicity,
we have also removed an overall $\delta$-function for the conservation of total
four-momentum from $T(p_1^\prime, p_2^\prime;p_1, p_2)$, 
$K(p_1^\prime, p_2^\prime;p_1, p_2)$ and $\Delta(k^2)$.  For example, 
$T(p_1^\prime, p_2^\prime;p_1, p_2)$ is defined by:
\begin{eqnarray}
(2\pi)^4 \delta^4(p_1^\prime + p_2^\prime - p_1 - p_2) T(p_1^\prime, p_2^\prime;p_1, p_2)
        &=& \nonumber \\
        && \hskip -0.5in
       \prod_i \bigl\{ ({p_i^\prime}^2 - m^2)(p_i^2 - m^2)
             \int d^4 x^\prime_i e^{ipx^\prime_i}\int d^4 x_i e^{-ipx_i} \bigr\} 
       \nonumber \\
   && \hskip -0.5in \cdot
\langle 0| \phi(x^\prime_1)\phi(x^\prime_2)\phi(x_1)\phi(x_1)|0 \rangle_{\rm conn}
\end{eqnarray}
where $\phi(x)$ is the quantized scalar field used in this example (normalized
so that the single particle pole in the 2-point function has unit residue),
$\langle \ldots \rangle_{\rm conn}$ indicates that only connected diagrams are
to be included and we are following the conventions of Peskin and 
Schoeder~\cite{Peskin:1995ev}.

For simplicity we write the sum over the internal four-momentum $\bar k$ in 
Eq.~\ref{eq:b-s-rel} as an integral.  However, this equation applies equally 
well to a finite volume system if the spatial part of this continuous integral 
is replaced by an appropriate discrete sum.  Specifically, in finite volume the 
total momentum operator $\vec P_{\rm op}$ is conserved and, given periodic 
spatial boundary conditions, has components which are quantized in units of 
$2\pi/L$.  The single-particle three momenta, $\vec p$, $\vec p_1\,^\prime$, 
$\vec P/2 + \bar k$, {\it etc.} correspond to indices in a finite-volume Fourier 
series and are also quantized in units of $2\pi/L$.

Equation~\ref{eq:b-s-rel} is more general than needed, constraining the 
off-shell, two-particle scattering amplitude, a matrix acting on a space
of functions of eight momentum components.  We must specialize this 
equation, reducing the number of momentum variables from eight to seven.
The choice of this seven-dimensional restriction of Eq.~\ref{eq:b-s-rel} is 
the key step in the desired generalization to non-zero total 3-momentum, 
$\vec P \ne 0$.  Once that has been done, the resulting equation will 
still be quite different from our non-relativistic target, 
Eq.~\ref{eq:b-s-non-rel}.  However, the final steps connecting 
Eq.~\ref{eq:b-s-rel} with Eqs.~\ref{eq:b-s-non-rel} and \ref{eq:se_intgr} 
proceed in a fashion very similar to those in L\"uscher's original derivation:  
We extract from the second term in Eq.~\ref{eq:b-s-rel} particular pieces 
which are regular functions of the total energy $P_0$.  By a re-arrangement 
procedure, these terms are incorporated in a modified kernel 
$\widetilde K(p_1^\prime, p_2^\prime;p_1, p_2)$.   After this step, our re-arranged 
equation will have the same form as the non-relativistic Eqs.~\ref{eq:b-s-non-rel}
and \ref{eq:se_intgr}.

In contrast to the non-relativistic case, our new equation will have 
a volume dependent potential.  However, if we remain below the four-particle
threshold and have introduced into our re-arranged kernel $\widetilde K$ only regular 
factors, the difference between the finite and infinite volume kernels will 
vanish exponentially in the box size.  Similar exponentially small errors will 
come from the failure of the resulting potential to have a truly finite range.  
Thus, we will be assured that the quantization condition in Eq.~\ref{eq:quantize} 
will apply to the relativistic case with $\vec P \ne 0$ up to exponentially 
small corrections.

We begin by restricting the general Bethe-Salpeter equation of 
Eq.~\ref{eq:b-s-rel} to a carefully defined seven-dimensional momentum
surface, making it similar to the target Eq.~\ref{eq:b-s-non-rel}.  (As
discussed earlier, Eq.~\ref{eq:b-s-non-rel} can also be viewed as involving
functions of seven dimensions provided the trivial dependence on the
total four-momentum is included.)

First express the pairs of two particle momenta $p_1^\prime, p_2^\prime$ and 
$p_1, p_2$ using relative and total momenta:
\begin{eqnarray}
p_{1,2}        &=& P/2 \pm k \nonumber \\
p_{1,2}^\prime &=& P/2 \pm k^\prime.
\end{eqnarray}
In Eq.~\ref{eq:b-s-rel} we have already imposed four-momentum conservation, 
removing an over-all momentum conserving delta function from both $T$ and 
$K$.  Next we impose a further condition on the four-momenta $k$ 
and $k^\prime$ requiring that $k_0 = \beta k_\parallel$, 
where $\beta=|\vec P|/P_0$ and $k_\parallel$ is the spatial component of 
the four-vector $k$ in the direction of $\vec P$.  A similar condition 
defines a restricted value for $k^\prime$ and the integration 
variable $\bar k$ in Eq.~\ref{eq:b-s-rel}.  This condition is not 
immediately useful since, while we can consistently impose it on the external 
$k$ and $k^\prime$ variables in Eq.~\ref{eq:b-s-rel}, the integral (or sum) 
over $\bar k$ does not obey any such restriction.  Note, the restriction 
$k_0 = \beta k_\parallel$ can be applied equally well in finite or infinite
volume since, while $k_\parallel$ will be discrete for the finite volume 
case, the time component, $k_0$, is always continuous and can be chosen to
obey such a relation.

To make progress we must remove from the integral over $\bar k$ some
terms which are not singular as $P_0$ approaches an allowed two-particle
energy, $P_0 \rightarrow \omega_+ + \omega_-$, where the single particle
energies are given by $\omega_\pm = \sqrt{(\vec P/2 \pm \vec k)^2 + m^2}$.  
This can be done by generalizing a discussion in Ref.~\cite{Luscher:1986pf} 
and arguing that the portion of the product 
$\Delta((P/2+\bar k)^2)\Delta((P/2-\bar k)^2)$ which is a singular function
of $P_0$ comes from the product of the single-particle singularities in
each of the $\Delta$ factors.  This singular term, arising when these 
poles pinch the $\bar k_0$ contour, can be viewed as a distribution in
$\bar k_0$, allowing us to write this product as:
\begin{eqnarray}
\Delta((P/2+k)^2)\Delta((P/2-k)^2) 
              &=& \frac{-i\pi}{P^2/4+k^2 -m^2+i\epsilon}\delta(P\cdot k)+R(P,k)
                    \nonumber \\
              &\equiv& S(P,k) + R(P,k), 
                    \label{eq:sing_dcomp}
\end{eqnarray}
where the function $R(P,\bar k)$ is a regular function of $P_0$ in the 
interval below the four-pion threshold:
\begin{equation}
2\sqrt{\frac{\vec P^2}{4}+m^2} \le P_0 \le 4\sqrt{\frac{\vec P^2}{16}+m^2}
\end{equation}
and the singular part, $S(P,k)$, is defined by Eq.~\ref{eq:sing_dcomp}.
Equation~\ref{eq:sing_dcomp} is derived in the Appendix and is a 
generalization of L\"uscher's Eq.~3.16 from Ref.~\cite{Luscher:1986pf}
to the case of $\vec P \ne 0$.

The next step absorbs the contribution from the regular function
$R(P,k)$ as follows.  First rewrite Eq.~\ref{eq:b-s-rel} in a more 
symbolic form exploiting the decomposition in Eq.~\ref{eq:sing_dcomp}:
\begin{equation}
T = K + K(S+R)T. 
\end{equation}
The second term on the right-hand-side can then be moved to the left hand side:
\begin{equation}
(1-KR)T = K + KST. 
\end{equation}
Finally we divide by the factor $(1-KR)$ and define 
\begin{equation}
\widetilde K = \frac{1}{1-KR}K
\end{equation}
The resulting equation can then be written:
\begin{equation}
T(k^\prime;k) = \widetilde K(k^\prime;k)
               + \int \frac{d^4 \bar k}{(2\pi)^4} \widetilde K(k^\prime;\bar k)
                  \frac{-i\pi\delta(\bar k\cdot P)}{\frac{P^2}{4} + \bar k^2 -m^2} T(\bar k;k).
\label{eq:b-s-transf}
\end{equation}
Here we have replaced the variables $p_i$ and $p^\prime_i$ with the total and
relative four-momenta $P$, $k^\prime$ and $k$ and suppressed the variable $P$.

It is now easy to see that Eq.~\ref{eq:b-s-transf} has a form identical to the
original non-relativistic Lippmann-Schwinger equation, Eq.~\ref{eq:b-s-non-rel}.
First we observe that the delta function $\delta(\bar k\cdot P)$ forces the 
integration four-momentum $\bar k$ to obey our restriction:
\begin{eqnarray}
0 &=& \bar k \cdot P = \bar k_0 P_0 - \bar k_\parallel |\vec P| \quad {\rm or} \quad
\bar k_0 = \beta \bar k_\parallel. 
\end{eqnarray}
This permits us to impose this relation between the time and parallel components 
of the relative four-momenta everywhere in this equation, effectively reducing it
to a three-dimensional integral equation as is the case for the non-relativistic
problem.  (As in that case, this equation is diagonal in the total four momentum,
$P$, the remaining four of our seven-dimensional momentum variables.)

Second, we observe that the denominator has a non-relativistic form if
we rescale the axis parallel to $\vec P$ by a factor of 
$\gamma = 1/\sqrt{1-\beta^2}$:
\begin{eqnarray}
\frac{P^2}{4} + \bar k^2 -m^2 
   &=& \frac{P_0^2}{4} - \frac{\vec P^2}{4} + k_0^2 - \vec \bar k^2 -m^2
                                     \nonumber \\
   &=& \frac{P_0^2}{4}-\frac{\vec P^2}{4}-\frac{1}{\gamma^2}\bar k_\parallel^2 
                                -\vec \bar k_\perp^2 -m^2.
\end{eqnarray}
Thus, if we change variables from $\bar k_\parallel$ to 
\begin{equation}
\widetilde k_\parallel = \frac{1}{\gamma}\bar k_\parallel,
\label{eq:contract}
\end{equation}
the denominator has the normal Laplacian form.  We obtain a complete match 
between the denominators in Eqs.~\ref{eq:b-s-non-rel} and \ref{eq:se_intgr}
and that in Eq.~\ref{eq:b-s-transf} if we remove a factor of $m$ 
from the denominator of Eq.~\ref{eq:b-s-transf} and identify the 
non-relativistic energy $E^{\rm cm}$ with $(P_0^2-\vec P^2)/4m - m$.

Here we should recall the standard connection between Bethe-Salpeter 
equation, Eq.~\ref{eq:b-s-rel} and the version of the Schr\"odinger
equation given by the homogenous Eq.~\ref{eq:se_intgr}, which determines
the discrete, finite-volume energies.  Since the Bethe-Salpeter equation,
{\it e.g.} Eq.~\ref{eq:b-s-transf}, holds in finite volume, the energy 
eigenvalues, $E_n$ for the finite-volume, relativistic system will correspond 
to poles in the 2-particle scattering amplitude $T$ obeying that equation.  
As one approaches the singularity at $P_0 \rightarrow E_n$ the inhomogeneous 
term, proportional to the kernel $K$, is not singular and can be dropped 
from the equation leaving a homogenous equation identical in form to 
Eq.~\ref{eq:se_intgr}.

Finally we must investigate the rotational symmetry of the kernel, 
$\widetilde K(k^\prime;k)$.  Since this function also depends on the four-vector 
$P$, there is possible rotationally asymmetric dependence on arguments of 
the form $\vec P \cdot \vec k^\prime$ and  $\vec P \cdot \vec k$ in addition 
to the acceptable dependence on $(\vec k^\prime)^2$, $\vec k^2$ and 
$\vec k^\prime \cdot k$.  Fortunately, if the components of $\vec k^\prime$ 
and $\vec k$ parallel to $\vec P$ are rescaled as described in 
Eq.~\ref{eq:contract}, it is easy to see that the resulting function 
$\widetilde K(k^\prime;k)$ becomes rotationally symmetric.  This can be 
demonstrated by exploiting the Lorentz invariance of the function
$\widetilde K(k^\prime;k)$ which is assured by the covariant separation of 
regular and singular parts used in Eq.~\ref{eq:sing_dcomp}.  We use this
Lorentz symmetry to equate $\widetilde K(k^\prime;k)$ to its value in the rest 
system:
\begin{equation}
\widetilde K(P;k^\prime;k) = \widetilde K(P_{cm};k_{cm}^\prime;k_{cm}),
\label{eq:transform}
\end{equation}
where for clarity we now also display the dependence on the total 
four-momentum $P$.

Since in the center of mass system $\vec P_{cm} = 0$, the right hand side of 
Eq.~\ref{eq:transform} is a manifestly rotationally invariant function of the center
of mass variables $\vec k^\prime_{cm}$ and $\vec k_{cm}$.  However we can easily 
express these variable in terms of the original $\vec k^\prime$ and $\vec k$.  For
example:
\begin{eqnarray}
\vec (k_{cm})_\perp &=& \vec k_\perp \label{eq:k_perp} \\
(k_{cm})_\parallel  &=& \gamma(k_\parallel - \beta k_0) = \frac{1}{\gamma}k_\parallel 
                         \label{eq:k_par} \\
(k_{cm})_0          &=& \gamma(k_0 - \beta k_\parallel) = 0,
\end{eqnarray}
where the second and third lines follow from the constraint obeyed by $k_0$, 
imposed when we specialized to this three-dimensional equation.  The third 
equation for $(k_{cm})_0$ is necessary to insure that rotational asymmetry is 
not introduced through a non-symmetric dependence of this variable on the 
original laboratory variables.

Note that the rescaled variables on which $\widetilde K(P;k^\prime;k)$ depends 
symmetrically, $(k_{cm})_\parallel = (1/\gamma)k_\parallel$ and 
$(k_{cm}^\prime)_\parallel = (1/\gamma) k^\prime_\parallel$, are precisely the variables 
in terms of which the denominator in Eq.~\ref{eq:b-s-transf} contains the standard
Laplacian.

For clarity, we now rewrite the resulting three-dimensional integral equation,
Eq.~\ref{eq:b-s-transf}, in terms of these translated momentum variables, labeled 
suggestively $\vec k_{\rm cm}$:
\begin{equation}
\langle \vec k_{\rm cm}^\prime| T |\vec k_{\rm cm}\rangle 
          = \langle  \vec k_{\rm cm}^\prime|\widetilde K|\vec k_{\rm cm}\rangle
   + \int \frac{d^3 \bar k_{\rm cm}}{(2\pi)^3 } 
             \langle  \vec k_{\rm cm}^\prime|\widetilde K|\vec{\bar{k}}_{\rm cm}\rangle
             \frac{-i}{2 P_0}\frac{1}{(P^2/4-m^2-\vec{\bar{k}}_{\rm cm}^2)}
             \langle  \vec{\bar{k}}_{\rm cm}|T|\vec k_{\rm cm}\rangle.
\label{eq:b-s-cm}
\end{equation}
While the variables $\vec k_{\rm cm}^\prime$, $\vec{\bar{k}}_{\rm cm}$ and 
$\vec k_{\rm cm}$ may appear to be Lorentz transformed, center-of-mass variables, 
in fact, they are simply the original variables defined in the laboratory system
except for a transformation of scale for the single component parallel to $\vec P$.  
This rescaling is equally well defined if the corresponding variable is continuous 
or discrete.  The only explicit use of Lorentz symmetry is to constrain the 
possible dependence of the function $\widetilde K(P,k^\prime,k)$ on its arguments.

Thus, we have demonstrated that our original, field-theoretic Bethe-Salpeter equation
can be rewritten as a non-relativistic Lippmann-Schwinger equation if we change to
rescaled variables $(k_{cm})_\parallel = (1/\gamma)k_\parallel$.  We must now ask
if such a rescaled set of momenta corresponds to an actual finite volume problem.
Recall that we began by examining a two-particle problem in a finite, cubic box of 
side $L$.  The transformation to the coordinates $\vec R$ and $\vec r$ of 
Eqs.~\ref{eq:rel_coord1} and \ref{eq:rel_coord2} and the choice of total 
momenta $\vec P$ then requires that the relative momenta $\vec k$ or $\vec k^\prime$ 
that appear in the Bethe-Salpeter Eq.~\ref{eq:b-s-transf} have the form 
$2\pi(n_1,n_2,n_3)/L$ where $n_i$ is an integer or half-integer depending on whether 
$LP_i/2\pi$ is an even or odd integer.  Now, we have recognized that this problem 
is equivalent to a non-relativistic problem with transformed momenta given by 
Eqs.~\ref{eq:k_perp} and \ref{eq:k_par}.  If these $k^{\rm cm}$ momenta correspond 
to those for a finite volume problem, then we have succeeded in casting the original 
relativistic problem into a non-relativistic problem which can be solved using the 
techniques developed by L\"uscher in Refs.~\cite{Luscher:1986pf,Luscher:1990ux}.

That this is in fact the case can be seen by examining two simple examples.  In 
the first example $\vec P = 2\pi(0,0,1)/L$.  In this case
the transformed variables $k^{\rm cm}_i$ take on a very simple form:
\begin{eqnarray}
k^{\rm cm}_1 &=& k_1 = \frac{2\pi}{L}n_1 \label{eq:qtm_ex_1a}\\
k^{\rm cm}_2 &=& k_2 = \frac{2\pi}{L}n_2 \label{eq:qtm_ex_1b} \\
k^{\rm cm}_3 &=& \gamma(k_3 - \beta k_0) = \frac{1}{\gamma}k_3  = \frac{2\pi}{\gamma L}n_3
             \label{eq:qtm_ex_1c}
\end{eqnarray}
where $n_1$, $n_2$ and $n_3+\frac{1}{2}$ are integers.  These quantized momenta 
correspond to a simple finite volume of length $L$ in the 1-and 2-directions and
expanded length $\gamma L$ in the 3-direction.  If periodic boundary conditions
are imposed in the 1- and 2-directions and anti-periodic conditions imposed in
the 3-direction, the resulting quantized momenta will correspond precisely with
those in Eqs.~\ref{eq:qtm_ex_1a}-\ref{eq:qtm_ex_1c}.  Thus, after generalizing
L\"uscher's non-relativistic technique to this sort of asymmetric box, we will
obtain the desired relation between the infinite-volume scattering phase shifts and
the discrete finite volume energies in this asymmetric box.  Of course, our analysis
above has determined that these same energies will be found in the original 
relativistic problem with total momentum $\vec P = 2\pi(0,0,1)/L$ and cubic box of 
side $L$.  This agrees with the result obtained Rummukainen and 
Gottlieb~\cite{Rummukainen:1995vs} and corresponds to a case of immediate practical
interest.

Now let us examine a second case where $\vec P = (1,1,0)2\pi/L$.  Using 
Eqs.~\ref{eq:k_perp} and \ref{eq:k_par} the rescaled momentum $\vec k^{\rm cm}$ is 
given by 
\begin{equation}
\vec k^{\rm cm} = \vec k - \frac{\vec k \cdot \vec P}{|\vec P|^2} \vec P
                  +\frac{1}{\gamma}\frac{\vec k \cdot \vec P}{|\vec P|^2}\vec P
\end{equation}
where the first two terms correspond to the untransformed perpendicular component
while the third term is the transformed parallel piece.  Written in terms of 
the individual components this equation becomes:
\begin{eqnarray}
k^{\rm cm}_1 &=& \Bigl\{\frac{n_1-n_2}{2} + \frac{n_1+n_2}{2\gamma}\Bigr\}\frac{2\pi}{L} 
                 \label{eq:qtm_ex_2a} \\
k^{\rm cm}_2 &=& \Bigl\{\frac{n_2-n_1}{2} + \frac{n_1+n_2}{2\gamma}\Bigr\}\frac{2\pi}{L} 
                 \label{eq:qtm_ex_2b} \\
k^{\rm cm}_3 &=& k_3 = n_3 \frac{2\pi}{L}\label{eq:qtm_ex_2c}.
\end{eqnarray}
These quantization conditions can be easily realized if we impose the following \\
(anti-)periodicity conditions on the wave function $\psi^{\rm cm}(\vec r)$ on
which the operators in the integral equation, Eq.~\ref{eq:b-s-cm}, act:
\begin{eqnarray}
\psi^{\rm cm}(\vec r) &=& - \psi^{\rm cm}(\vec r + \vec D_i), \quad \mbox{\rm for}\; i=1,2 
                           \label{eq:quant_perp} \\
\psi^{\rm cm}(\vec r) &=& + \psi^{\rm cm}(\vec r + \hat e_3 L )
                           \label{eq:quant_par}
\end{eqnarray}
where the displacement vector $D_i$ is chosen to pick out the integer $n_i$
from the dot product $\vec D_i \cdot \vec k^{\rm cm} = 2\pi n_i L$:
\begin{eqnarray}
D_1 = \hat e_1 \frac{\gamma +1}{2}L + \hat e_2 \frac{\gamma-1}{2}L \\
D_2 = \hat e_1 \frac{\gamma -1}{2}L + \hat e_2 \frac{\gamma+1}{2}L.
\end{eqnarray}

The quantization condition given in Eqs.~\ref{eq:quant_perp} and \ref{eq:quant_par}
is equivalent to requiring that the wavefunction $\psi^{\rm cm}(\vec r)$ obey
anti-periodic boundary conditions on the faces of a rhombus whose sides are parallel
to the vectors $D_i$ and whose diagonals have length $|D_1+D_2|=\sqrt{2}\gamma L$
and $|D_1-D_2|=\sqrt{2} L$, precisely the earlier result of Rummukainen and 
Gottlieb~\cite{Rummukainen:1995vs}.  From these two examples, it is clear that
the case of general total momentum $\vec P = (n_1,n_2,n_3) 2\pi/L$ can also be 
realized by imposing (anti-)periodic on an appropriately distorted volume.

The results of this section can be summarized by returning to our first 
example, the case of a symmetrical $L^3$ box and a two-pion state with total 
momentum oriented in the 3-direction: $\vec P = (2\pi/L)\hat e_3$.  
Under these circumstances, the Bethe-Salpeter equation obeyed by the 
two-particle, off-shell scattering amplitude has been shown to be equivalent 
to a Schr\"odinger-like wave equation obeyed in an asymmetrical box with 
sides $L \times L \times \gamma L$ with the longer $\gamma L$ side parallel 
to the 3-direction.  Thus, up to exponentially small corrections, the 
energy eigenvalues of the original 2-particle, $L^3$ system can be predicted 
using this $L \times L \times \gamma L$, Schr\"odinger-like system.  As a
result, the allowed energies $E$ of this original system must obey a 
quantization condition similar to that given in Eq.~\ref{eq:quantize} 
where the function $\phi(k)$ in that equation must be modified to describe 
the anti-periodic boundary conditions and expanded length in the 3-direction:
\begin{eqnarray}
n\pi-\delta_0(k_{cm})  &=& \phi(q) \label{eq:quantize2}\\
\tan \phi(q)         &=& -\frac{\gamma \pi^{3/2} q}{{\cal Z}_{00}(1;q^2;\gamma)}, 
                                                    \nonumber \\
\phi(0)              &=& 0 \label{eq:new-phi1}\\
{\cal Z}_{00}(s;q^2;\gamma) &=& \frac{1}{\sqrt{4\pi}} 
                    \sum_{n \in {\mathbb Z}^3}
                       (n_1^2+n_2^2+\frac{1}{\gamma^2}(n_3+\frac{1}{2})^2-q^2)^{-s},
                           \label{eq:new-phi2}
\end{eqnarray}
where $n$ is an integer, $k_{cm} = \frac{1}{2}\sqrt{P^2-(2m)^2}$ and $q = k_{cm} L/2\pi$.  
This is the original result of Rummukainen and Gottlieb~\cite{Rummukainen:1995vs}.

\section{Two-particle decay of states with non-zero momentum}
\label{sec:decay}

The last part of this discussion is a generalization of the arguments of 
Lellouch and L\"uscher in Ref.~\cite{Lellouch:2000pv} to the case of non-zero
total momenta.  Fortunately, this is very straight forward because the methods
employed in that paper work equally well for $\vec P \ne 0$ once the formula
relating energy levels and scattering phase shifts has been generalized to
this non-zero momentum case.

We begin by reviewing the Lellouch-L\"uscher approach, which is somewhat 
indirect.  For our present purposes we will describe this method for the 
case of non-zero center-of-mass momentum, $\vec P \ne 0$.  One considers 
a finite volume system with both a $K^0$ meson and a degenerate $\pi^+\pi^-$ 
state, each with total four-momentum $P$.  Here the finite volume has been 
adjusted to insure that $E_{\pi\pi}=\sqrt{m_K^2 + \vec P^2}$, including the 
effects of the $\pi-\pi$ interaction.  Next, the effects of the weak 
interaction Hamiltonian, $H_W$, mixing these two states, are then
examined in perturbation theory.  To zeroth order, the $K^0$ and 
$\pi^+\pi^-$ states are degenerate and non-interacting.  To first order 
these states mix and their energies can be determined in first order, 
degenerate perturbation theory: 
\begin{eqnarray}
P_0 &=&           \sqrt{m_K^2 + \vec P^2}                 \label{eq:decay_fv_1} \\ 
    &\rightarrow& P_0 + \Delta P_0                        \label{eq:decay_fv_2} \\
    &=&           P_0 \pm\langle\pi^+\pi^-|H_W|K^0\rangle \label{eq:decay_fv_3} \\
    &\equiv&      P_0 \pm M,                              \label{eq:decay_fv_4}
\end{eqnarray}
where the states appearing in this formula are finite-volume states with
non-zero total momentum, normalized to unit probability.

The next step relates the finite volume amplitude $M$, which can be 
computed directly in a lattice QCD calculation, with the infinite
volume matrix element of $H_W$ that determines the physical partial 
width. This step uses Eq.~\ref{eq:quantize2} to relate the infinite 
volume phase shift, computed at the quantized, finite-volume energy 
$P_0+\Delta P_0$ with that energy shift related to $M$ by 
Eqs.~\ref{eq:decay_fv_2}-\ref{eq:decay_fv_4}.  Since the effect of 
$H_W$ on the scattering phase shift can be determined analytically in 
terms of the infinite-volume matrix elements of $H_W$, this equation 
will then relate the known, finite volume matrix element of $H_W$ 
with the desired, infinite volume matrix element.

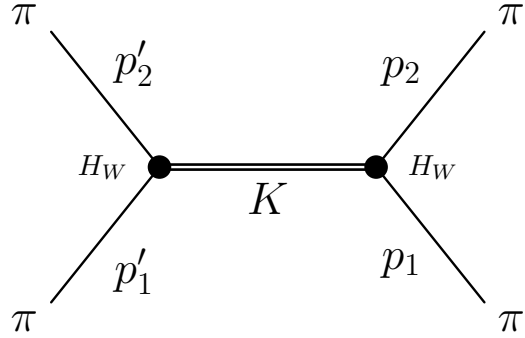
\begin{figure}
\centering
\begin{fmffile}{kres}
  \begin{fmfchar*}(60,30)
    \fmf{vanilla,label={\Large $p_1$},label.side=left}{p1in,kpp1}
    \fmf{vanilla,label={\Large $p_2$},label.side=right}{p2in,kpp1}
    \fmflabel{\Large $\pi$}{p1in} 
    \fmflabel{\Large $\pi$}{p2in}
    \fmfright{p1in,p2in}
    \fmfleft{p1out,p2out}
    \fmf{double,label={\Large $K$},label.side=left}{kpp1,kpp2}
    \fmflabel{\Large $\pi$}{p1out} 
    \fmflabel{\Large $\pi$}{p2out}
    \fmf{vanilla,label={\Large $p_1^\prime$},label.side=left}{kpp2,p1out}
    \fmf{vanilla,label={\Large $p_2^\prime$},label.side=right}{kpp2,p2out}
    \fmflabel{$H_W$}{kpp1}
    \fmfv{label.angle=0,label.dist=12,decor.shape=circle,decor.filled=full,decor.size=4thick}{kpp1}
    \fmflabel{$H_W$}{kpp2}
    \fmfv{label.angle=180,label.dist=12,decor.shape=circle,decor.filled=full,decor.size=4thick}{kpp2}
  \end{fmfchar*}
\end{fmffile}

\caption{The contribution of $H_W$ to $\pi-\pi$ scattering involving resonant production 
of a $K$ meson.  Because of the singular $K$-meson propagator, the amplitude corresponding
to this graph will be first order in $H_W$ when evaluated at the center-of-mass energies 
$m_K \pm \gamma |M|$ as required for our application.}
\label{fig:K_res}
\end{figure}

Thus, we must compute the variation in the $\pi-\pi$ scattering phase 
shift caused by the resonant scattering into the $K$ meson state.  
This infinite volume calculation is most easily done in the $\pi-\pi$ 
center of mass system and follows easily from the single $\pi-\pi$ 
scattering diagram with a $K$ meson intermediate state shown in 
Fig.~\ref{fig:K_res} giving:
\begin{equation}
\Delta\delta_0(k_{cm}) = - \frac{k_{cm} |A|^2}{32\pi m_K^2 \Delta (P_0)_{cm}}.
\label{eq:delta_delta}
\end{equation}
Here we have evaluated the addition to the $\pi-\pi$ scattering phase 
shift coming from the resonant scattering into the $K$ meson state at a
center-of-mass energy which corresponds to the laboratory energy
$P_0+\Delta P_0$ determined by Eqs.~\ref{eq:decay_fv_4}:
\begin{equation}
\Delta (P_0)_{cm} = \frac{\partial \sqrt{P_0^2 - \vec P^2}}{\partial P_0} \Delta P_0
                  = \gamma \Delta P_0 = \pm \gamma |M|
\end{equation}

The infinite volume decay amplitude $A$ appearing in Eq.~\ref{eq:delta_delta} 
is normalized following the conventions of Lellouch and L\"uscher so that 
the corresponding decay width is given by:
\begin{equation}
\Gamma_{K \rightarrow \pi\pi}  = \frac{k_{cm}}{16\pi m_K^2} |A|^2.
\end{equation}

Finally, these results are used to evaluate the terms in the quantization
condition, Eq.~\ref{eq:quantize2} for the original finite volume 
$\vec P \ne 0$, $\pi-\pi$ system.  The terms in this equation which are 
first order in the matrix elements of $H_W$ are:
\begin{equation}
-\Delta k_{cm} \Big\{\frac{\partial \delta_0(k)}{\partial k} \Big\}_{k=k_{cm}} 
            + \frac{k_{cm} |A|^2}{32\pi m_K^2 \Delta (P_{cm})_0} 
     = \Delta k_{cm} \Big\{\frac{\partial \phi(q)}{\partial k} \Big\}_{k=k_{cm}}
\end{equation}
where relation between $\Delta k_{cm}$ and $\Delta P_0$ is determined by
$k_{cm} = \sqrt{P^2/4-m^2}$ which implies:
\begin{equation}
\Delta k_{cm} = \frac{P_0 \Delta P_0}{4 k_{cm}}.
\label{eq:decay_iv}
\end{equation}
Combining Eqs.~\ref{eq:decay_fv_4} and \ref{eq:decay_iv} then gives the
desired connection between $A$ and \\ $M = \langle\pi^+\pi^-|H_W|K^0\rangle$:
\begin{equation}
\mp\frac{P_0 |M|}{4k_{cm}} \Big\{\frac{\partial \delta_0(k)}{\partial k} \Big\}_{k=k_{cm}} 
            \pm \frac{k_{cm} |A|^2}{32\pi m_K^2 \gamma |M|} 
     = \pm\frac{P_0 |M|}{4k_{cm}} \Big\{\frac{\partial \phi(q)}{\partial k} \Big\}_{k=k_{cm}}.
\end{equation}
If this equation is solved for $|A|^2$ we obtain the desired generalization of
the original Lellouch-L\"uscher condition to the case of $\vec P \ne 0$:
\begin{equation}
|A|^2 = 8\pi\frac{m_K^3}{k_{cm}^3} \gamma^2 
                      \Bigl\{k\frac{\partial \delta}{\partial k}  
                           + q\frac{\partial \phi}{\partial q} \Bigr\}|M|^2.
\end{equation}
This formula differs from that of Ref.~\cite{Lellouch:2000pv}
by the presence of the factor of $\gamma^2$ and the more complex, 
$\gamma$-dependent definition of the function $\phi(q)$ given in 
Eqs.~\ref{eq:new-phi1} and \ref{eq:new-phi2}.

\section{Conclusion}

In the preceding sections we have examined the case of two interacting
particles confined in a finite spatial volume and carrying non-zero 
total momentum.  We have determined a relation between the quantized 
energies of these finite-volume states and the two-particle scattering 
phase shifts.  This result, first obtained by Rummukainen and Gottlieb 
in Ref.~\cite{Rummukainen:1995vs}, is here obtained from the Bethe-Salpeter 
equation of relativistic quantum field theory using an extension of the 
methods that L\"uscher applied to the case of zero total momentum in 
Refs.~\cite{Luscher:1986pf,Luscher:1990ux}.  We then exploit these finite 
volume results to analyze two-particle decays.  The result, an extension of 
earlier work of Lellouch and L\"uscher, Ref.~\cite{Lellouch:2000pv}, to
the case of non-zero total momentum, provides an explicit formula that 
relates finite volume decay matrix elements computed using lattice gauge 
theory techniques and the infinite-volume quantities that enter physical 
decay rates.  

The ability to work with states with non-zero total momentum when 
computing such decay matrix elements offers two potentially important 
benefits to the study of $K \rightarrow \pi \pi$ decay.  First, by a 
proper choice of the total momentum, the corresponding two-pion state 
with lowest energy can be arranged to have an energy equal to that of the 
K meson, permitting the direct calculation of a physical, on-shell, decay 
matrix element.  Second, by working with a K meson with non-zero momentum, 
we insure that the unphysical vacuum decay amplitude, normally dominant 
in such a Euclidean matrix element calculation, will vanish because of 
momentum conservation.  We are now exploring this approach numerically.

\appendix
\section{Singular part of two-propagator product}

The singular contribution to the integral in Eq.~\ref{eq:b-s-rel} will
come as the energy $P_0$ is adjusted to cause singularities present in
both of the single-particle propagator factors, $\Delta((P/2\pm \bar k)^2)$, 
to pinch the $\bar k_0$ contour.  Thus, we can obtain the singular part 
of the integral over $\bar k_0$ in Eq.~\ref{eq:b-s-rel} by replacing both 
propagators by their singular part:
\begin{equation}
\Delta((P/2\pm k)^2) \rightarrow \frac{i}{(P/2\pm k)^2 - m^2}.
\end{equation}
Equation~\ref{eq:sing_dcomp} is based on the following expression
for the singular part of the product of two free scalar propagators:
\begin{equation}
\frac{i}{(P/2+k)^2-m^2 + i \epsilon} \frac{i}{(P/2-k)^2-m^2 + i \epsilon}
             \equiv \frac{-i\pi\delta(P\cdot k)}{P^2/4+k^2 -m^2+i\epsilon},
                 \label{eq:sing_dcomp2}
\end{equation}
where these two quantities are equivalent in the sense that they have
the same pole in the total energy $P_0$ at the point 
$P_0=\omega_+ + \omega_-$

Following L\"uscher, this formula is interpreted as relating two 
distributions in the variable $k_0$ over the space of test functions
$f(k_0)$ analytic in $k_0$ within a band around the imaginary $k_0$
axis.  Their equivalence can be demonstrated by multiplying by such 
a test function and then integrating $k_0$ along the imaginary axis.  
The left-hand-side of Eq.~\ref{eq:sing_dcomp2} is evaluated by moving 
the $k_0$ contour past one of the two pinching poles, 
$k_0 = P_0/2 - \omega_-$ or $k_0 = -P_0/2 + \omega_+$ and keeping the 
contribution of that pole given by Cauchy's theorem.  Following this 
procedure the left-hand-side of Eq.~\ref{eq:sing_dcomp2} becomes:
\begin{equation}
-i\pi\frac{f((\omega_+ - \omega_-)/2)}{2\omega_+\omega_-(P_0-\omega_+-\omega_-)}
\label{eq:lhs}
\end{equation}
where we have simplified this expression by replacing $P_0$ in the 
residue of the $P_0 = \omega_+ + \omega_-$ pole by its value at that pole.
Next we evaluate the right-hand-side of Eq.~\ref{eq:sing_dcomp2}
by using the delta-function to evaluate the integral over $k_0$.  We obtain:
\begin{equation}
-i\pi \frac{f(\frac{\vec P \cdot \vec k}{P_0})}
           {P_0\Bigl\{P^2_0/4 + (\frac{\vec P \cdot \vec k}{P_0})^2 
                 - {\vec P}^2/4 - {\vec k}^2 -m^2 +i\epsilon}\Bigr\}
\label{eq:rhs}
\end{equation}
where for clarity the $P_0$ arguments have not been simplified.  The final 
step requires computing the location and residue of the pole in $P_0$ 
found in the expression in Eq.~\ref{eq:rhs}.  The result agrees precisely 
with that shown in Eq.~\ref{eq:lhs} provided one is careful to include 
the effects of the $(\vec P \cdot \vec k/P_0)^2$ term when computing the 
residue and then replaces $P_0$ in the residue with $\omega_+ + \omega_-$, 
it value at the pole.  We forego a discussion of the specific region of 
analyticity for the test functions above and its consistency with the 
analytic properties of the amplitudes in the Bethe-Salpeter equation, 
Eq.~\ref{eq:b-s-rel}, since these kinematics should be a direct Lorentz 
transform of those discussed by L\"uscher in Ref.~\cite{Luscher:1986pf}.

\bibliography{paper}

\begin{thebibliography}{1}
\expandafter\ifx\csname bibnamefont\endcsname\relax
  \def\bibnamefont#1{#1}\fi
\expandafter\ifx\csname bibfnamefont\endcsname\relax
  \def\bibfnamefont#1{#1}\fi
\expandafter\ifx\csname url\endcsname\relax
  \def\url#1{\texttt{#1}}\fi
\expandafter\ifx\csname urlprefix\endcsname\relax\def\urlprefix{URL }\fi
\expandafter\ifx\csname bibinfo\endcsname\relax \def\bibinfo#1#2{#2}\fi
\expandafter\ifx\csname eprint\endcsname\relax \def\eprint#1{#1}\fi

\bibitem{Maiani:1990ca}
\bibinfo{author}{\bibfnamefont{L.}~\bibnamefont{Maiani}} \bibnamefont{and}
  \bibinfo{author}{\bibfnamefont{M.}~\bibnamefont{Testa}},
  \bibinfo{journal}{Phys. Lett.} \textbf{\bibinfo{volume}{B245}},
  \bibinfo{pages}{585} (\bibinfo{year}{1990}).

\bibitem{Lellouch:2000pv}
\bibinfo{author}{\bibfnamefont{L.}~\bibnamefont{Lellouch}} \bibnamefont{and}
  \bibinfo{author}{\bibfnamefont{M.}~\bibnamefont{Luscher}},
  \bibinfo{journal}{Commun. Math. Phys.} \textbf{\bibinfo{volume}{219}},
  \bibinfo{pages}{31} (\bibinfo{year}{2001}), \eprint{hep-lat/0003023}.

\bibitem{Lin:2001ek}
\bibinfo{author}{\bibfnamefont{C.~J.~D.} \bibnamefont{Lin}},
  \bibinfo{author}{\bibfnamefont{G.}~\bibnamefont{Martinelli}},
  \bibinfo{author}{\bibfnamefont{C.~T.} \bibnamefont{Sachrajda}},
  \bibnamefont{and} \bibinfo{author}{\bibfnamefont{M.}~\bibnamefont{Testa}},
  \bibinfo{journal}{Nucl. Phys.} \textbf{\bibinfo{volume}{B619}},
  \bibinfo{pages}{467} (\bibinfo{year}{2001}), \eprint{hep-lat/0104006}.

\bibitem{Luscher:1986pf}
\bibinfo{author}{\bibfnamefont{M.}~\bibnamefont{Luscher}},
  \bibinfo{journal}{Commun. Math. Phys.} \textbf{\bibinfo{volume}{105}},
  \bibinfo{pages}{153} (\bibinfo{year}{1986}).

\bibitem{Luscher:1990ux}
\bibinfo{author}{\bibfnamefont{M.}~\bibnamefont{Luscher}},
  \bibinfo{journal}{Nucl. Phys.} \textbf{\bibinfo{volume}{B354}},
  \bibinfo{pages}{531} (\bibinfo{year}{1991}).

\bibitem{Rummukainen:1995vs}
\bibinfo{author}{\bibfnamefont{K.}~\bibnamefont{Rummukainen}} \bibnamefont{and}
  \bibinfo{author}{\bibfnamefont{S.~A.} \bibnamefont{Gottlieb}},
  \bibinfo{journal}{Nucl. Phys.} \textbf{\bibinfo{volume}{B450}},
  \bibinfo{pages}{397} (\bibinfo{year}{1995}), \eprint{hep-lat/9503028}.

\bibitem{kss}
\bibinfo{author}{\bibfnamefont{C.}~\bibnamefont{Kim},
  \bibfnamefont{C.h.~Sachrajda}} \bibnamefont{and}
  \bibinfo{author}{\bibfnamefont{S.}~\bibnamefont{Sharpe}},
  \bibinfo{journal}{Southampton, Univ. of Washington preprint}
  (\bibinfo{year}{2005}).

\bibitem{Peskin:1995ev}
\bibinfo{author}{\bibfnamefont{M.~E.} \bibnamefont{Peskin}} \bibnamefont{and}
  \bibinfo{author}{\bibfnamefont{D.~V.} \bibnamefont{Schroeder}}
  \bibinfo{note}{Reading, USA: Addison-Wesley (1995) 842 p}.

\end{thebibliography}

\end{document}